\documentclass[apj]{emulateapj}
\usepackage{psfig,amsfonts,amsmath,graphicx,natbib,apjfonts}
\def\ra#1#2#3{#1$^{\rm h}$#2$^{\rm m}$#3$^{\rm s}$}
\def\dec#1#2#3{#1$^\circ$#2$'$#3$''$}
\def\nod{\nodata}
\newcommand{\gps}{\ensuremath{g_{\rm P1}}}
\newcommand{\rps}{\ensuremath{r_{\rm P1}}}
\newcommand{\ips}{\ensuremath{i_{\rm P1}}}
\newcommand{\zps}{\ensuremath{z_{\rm P1}}}
\newcommand{\yps}{\ensuremath{y_{\rm P1}}}
\newcommand{\grizy}{\gps\rps\ips\zps\yps}

\def\cfa{1}
\def\ucsc{2}
\def\stsci{3}
\def\pri{4}
\def\ifa{5}
\def\dur{6}
\def\har{7}

\begin{document}

\title{A Search for Fast Optical Transients in the Pan-STARRS1
Medium-Deep Survey: M Dwarf Flares, Asteroids, Limits on Extragalactic
Rates, and Implications for LSST}

\author{
E.~Berger\altaffilmark{\cfa},
C.~N.~Leibler\altaffilmark{\cfa,}\altaffilmark{\ucsc},
R.~Chornock\altaffilmark{\cfa},
A.~Rest\altaffilmark{\stsci},
R.~J.~Foley\altaffilmark{\cfa},
A.~M.~Soderberg\altaffilmark{\cfa},
P.~A.~Price\altaffilmark{\pri},
W.~S.~Burgett\altaffilmark{\ifa},
K.~C.~Chambers\altaffilmark{\ifa},
H.~Flewelling\altaffilmark{\ifa},
M.~E.~Huber\altaffilmark{\ifa},
E.~A.~Magnier\altaffilmark{\ifa},
N.~Metcalfe\altaffilmark{\dur},
C.~W.~Stubbs\altaffilmark{\har},
\& J.~L.~Tonry\altaffilmark{\ifa}
}

\altaffiltext{1}{Harvard-Smithsonian Center for Astrophysics, 60
Garden Street, Cambridge, MA 02138, USA}

\altaffiltext{2}{Department of Astronomy and Astrophysics, University
of California, Santa Cruz, CA 95064, USA}

\altaffiltext{3}{Space Telescope Science Institute, 3700 San Martin
Drive, Baltimore, Maryland 21218, USA}

\altaffiltext{4}{Department of Astrophysical Sciences, Princeton
University, Princeton, NJ 08544, USA}

\altaffiltext{5}{Institute for Astronomy, University of Hawaii, 2680
Woodlawn Drive, Honolulu, HI 96822, USA}

\altaffiltext{6}{Department of Physics, Durham University, South Road,
Durham DH1 3LE, UK}

\altaffiltext{7}{Department of Physics, Harvard University, Cambridge,
MA 02138, USA}

\begin{abstract} We present a search for fast optical transients
($\tau\sim 0.5\,{\rm hr}-1\,{\rm d}$) using repeated observations of
the Pan-STARRS1 Medium-Deep Survey (PS1/MDS) fields.  Our search takes
advantage of the consecutive \gps\rps\ observations ($16.5$ min in
each filter), by requiring detections in both bands, with
non-detections on preceding and subsequent nights.  We identify $19$
transients brighter than $22.5$ AB mag (${\rm S/N}\gtrsim 10$).  Of
these, 11 events exhibit quiescent counterparts in the deep PS1/MDS
templates that we identify as M4--M9 dwarfs at $d\approx 0.2-1.2$ kpc.
The remaining 8 transients lack quiescent counterparts, exhibit mild
but significant astrometric shifts between the \gps\ and \rps\ images,
colors of $(g-r)_{\rm P1} \approx 0.5-0.8$ mag, non-varying light
curves, and locations near the ecliptic plane with solar elongations
of about $130$ deg, which are all indicative of main-belt asteroids
near the stationary point of their orbits.  With identifications for
all 19 transients, we place an upper limit of $R_{\rm FOT}(\tau\sim
0.5\,{\rm hr})\lesssim 0.12$ deg$^{-2}$ d$^{-1}$ ($95\%$ confidence
level) on the sky-projected rate of extragalactic fast transients at
$\lesssim 22.5$ mag, a factor of $30-50$ times lower than previous
limits; the limit for a timescale of $\sim {\rm day}$ is $R_{\rm
FOT}\lesssim 2.4\times 10^{-3}$ deg$^{-2}$ d$^{-1}$.  To convert these
sky-projected rates to volumetric rates, we explore the expected peak
luminosities of fast optical transients powered by various mechanisms,
and find that non-relativistic events are limited to $M\approx -10$
mag ($M\approx -14$ mag) for a timescale of $\sim 0.5$ hr ($\sim {\rm
day}$), while relativistic sources (e.g., gamma-ray bursts,
magnetar-powered transients) can reach much larger luminosities.  The
resulting volumetric rates are $\lesssim 13$ Mpc$^{-3}$ yr$^{-1}$
($M\approx -10$ mag), $\lesssim 0.05$ Mpc$^{-3}$ yr$^{-1}$ ($M\approx
-14$ mag) and $\lesssim 10^{-6}$ Mpc$^{-3}$ yr$^{-1}$ ($M\approx -24$
mag), significantly above the nova, supernova, and GRB rates,
respectively, indicating that much larger surveys are required to
provide meaningful constraints.  Motivated by the results of our
search we discuss strategies for identifying fast optical transients
in the LSST main survey, and reach the optimistic conclusion that the
veil of foreground contaminants can be lifted with the survey data,
without the need for expensive follow-up observations.  \end{abstract}


\section{Introduction}
\label{Sec:Intro}

For nearly a century, optical observations aimed at the discovery and
study of astrophysical transients have largely focused on events with
durations of days to months.  This is mainly due to a fortuitous match
with the timescales of the most common extragalactic events (novae and
supernovae), whose typical luminosities and intrinsic rates require
coverage of large numbers of galaxies and/or blank sky areas, leading
to a natural search cadence of several days.  Thus, novae have a much
higher intrinsic rate than supernovae ($\sim 2.2$ yr$^{-1}$ per
$10^{10}$ L$_{\rm K,\odot}$ versus $\sim 2\times 10^{-3}$ yr$^{-1}$
per $10^{10}$ L$_{\rm K,\odot}$, respectively; e.g.,
\citealt{ws04,lcl+11}), but supernovae are significantly more luminous
than novae ($\sim -18$ mag versus $\sim -8$ mag, respectively;
\citealt{gs78,fil97}).  As a result, for a given survey limiting
magnitude tens of novae can be discovered per year by targeting a few
nearby galaxies with a cadence of few days, while discovering a
similar number of supernovae requires monitoring of $\sim 10^4$
galaxies (or hundreds of deg$^2$), thereby necessitating a similar
cadence of several days; for the purpose of sheer discovery rate, a
faster cadence is not profitable for nova and supernova searches.
Over the past few decades such surveys have been highly successful at
discovering about one hundred novae and supernovae per year (e.g.,
\citealt{ws04,llc+11}).

The advent of large format cameras on dedicated wide-field telescopes,
coupled with serendipitous discoveries of transients outside of the
traditional nova and supernova luminosity and timescale ranges, has
opened up a new discovery space for astrophysical transients.  By
repeatedly targeting the same fields, such surveys are in principle
capable of exploring a wide range of timescales, from the duration of
single exposures (i.e., minutes) to years.  In practice, most surveys
are still primarily focused on supernovae (driven to a large extent by
Type Ia supernova cosmology at increasingly larger redshifts), and
therefore cover wider fields to greater depth at the expense of a
faster temporal cadence to maximize the supernova discovery rate while
preserving adequate light curve coverage.  Still, some surveys have
been utilized to perform initial searches for fast optical transients
(FOTs) on timescales as short as $\sim 0.5$ hr.  Clearly, the
effective areal exposure of such searches (i.e., the product of survey
area and exposure time) becomes progressively smaller at faster
cadence as sky coverage has to be sacrificed for repeated
short-cadence observations.

In this context, the Deep Lens Survey (DLS) was utilized to search for
FOTs on a timescale of about $1300$ s to a depth of $B\approx 23.8$
mag (with a total exposure of 1.1 deg$^2$ d) and led to an upper limit
on the extragalactic sky-projected rate of $R_{\rm FOT}\lesssim 6.5$
deg$^{-2}$ d$^{-1}$ ($95\%$ confidence level; \citealt{bwb+04}).  The
DLS search uncovered three fast transients, which were shown to be
flares from Galactic M dwarf stars \citep{bwb+04,kr06}.  A search for
transients with a timescale of $\gtrsim 0.5$ hr to a depth of about
17.5 mag (with a total exposure of 635 deg$^2$ d) using the Robotic
Optical Transient Search Experiment-III (ROTSE-III) yielded a limit on
the extragalactic rate of $R_{\rm FOT}\lesssim 5\times 10^{-3}$
deg$^{-2}$ d$^{-1}$, and uncovered a single M dwarf flare
\citep{raa+05}; a similar search with the MASTER telescope system
yielded comparable limits \citep{lkk+07}, and two uncharacterized
candidate fast transients \citep{glk+13}.  Similarly, a targeted
search for transients on a timescale of about 0.5 hr and to a depth of
$B\approx 21.3$ mag in the Fornax galaxy cluster (with a total
exposure of 1.9 deg$^2$ d) placed a limit on the extragalactic rate of
$R_{\rm FOT}\lesssim 3.3$ deg$^{-2}$ d$^{-1}$ \citep{rok+08}.  Two
fast transients were detected in this search, both shown to be M dwarf
flares \citep{rok+08}.  At the bright end, the ``Pi of the Sky''
project placed a limit on transients brighter than 11 mag with a
duration of $\gtrsim 10$ s of $R_{\rm FOT}\lesssim 5\times 10^{-5}$
deg$^{-2}$ d$^{-1}$ \citep{smp+10}.

Fast optical transients have also been found serendipitously by other
surveys, but they have generally been shown to be Galactic in
origin\footnotemark\footnotetext{A recent example is a large amplitude
transient from the Catalina Real-Time Transient Survey, which was
initially claimed to be extragalactic in origin \citep{atel4586}, but
was subsequently shown to be an M dwarf flare \citep{atel4619}.
Additional cautionary tales are summarized by \citet{kr06}.}.  A
notable exception is the transient PTF11agg \citep{ckh+13}, which
faded by about 1.2 mag in 5.3 hours, and 3.9 mag in 2.2 d, and was
accompanied by radio emission that may be indicative of relativistic
expansion (although the distance of this transient is not known,
thereby complicating its interpretation).  We note that such an event,
while fast compared to the general nova and supernova population, is
still of longer duration than the timescales probed by the DLS and
Fornax searches, as well as the Pan-STARRS1 search we describe here.
Thus, the existing searches for extragalactic FOTs have mainly raised
the awareness that the foreground of M dwarf flares is large, with an
estimated all-sky rate of $\sim 10^8$ yr$^{-1}$ at a limiting
magnitude of $\sim 24$ mag \citep{bwb+04,kr06}.

Here, we present a search for fast optical transients with an
effective timescale of about 0.5 hr to 1 d and to a depth of about
22.5 mag in the first $1.5$ years of data from the Pan-STARRS1
Medium-Deep Survey (PS1/MDS).  This search uncovered a substantial
sample of 19 fast transients, both with and without quiescent
counterparts.  We describe the survey strategy and selection criteria
in \S\ref{sec:obs}.  In \S\ref{sec:results} we summarize the
properties of the 19 detected transients and classify them using a
combination of color information, astrometry, sky location, and the
properties of quiescent counterparts (when detected).  With a unique
identification of all 19 transients as Solar system or Galactic in
origin, we place a limit on the rate of extragalactic fast transients
that is $30-50$ times better than the limits from previous searches
(\S\ref{sec:rate}).  We further investigate for the first time the
limits on the {\it volumetric} rates of FOTs from our survey and
previous searches using the survey limiting magnitudes and fiducial
transients luminosities.  In \S\ref{sec:models} we expand on this
point and discuss the expected peak luminosities of FOTs for a range
of physically motivated models.  Finally, since our search is the
first one to utilize observations that are similar to the Large
Synoptic Survey Telescope (LSST) main survey strategy, we conclude by
drawing implications for fast optical transient searches in the LSST
data \S\ref{sec:lsst}.

\begin{figure*}
\centering
\includegraphics[angle=0,width=0.48\textwidth]{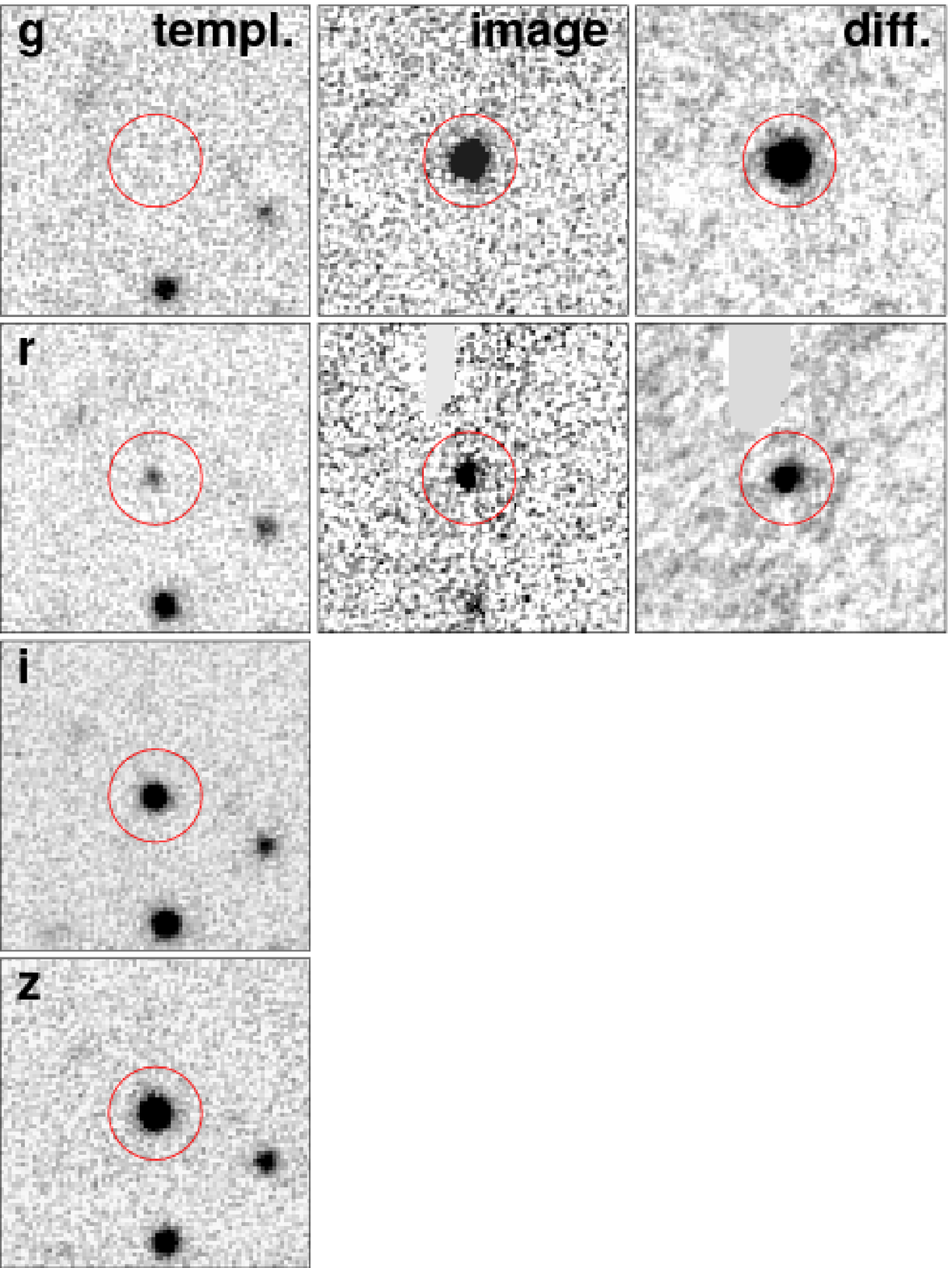}\hfill
\includegraphics[angle=0,width=0.48\textwidth]{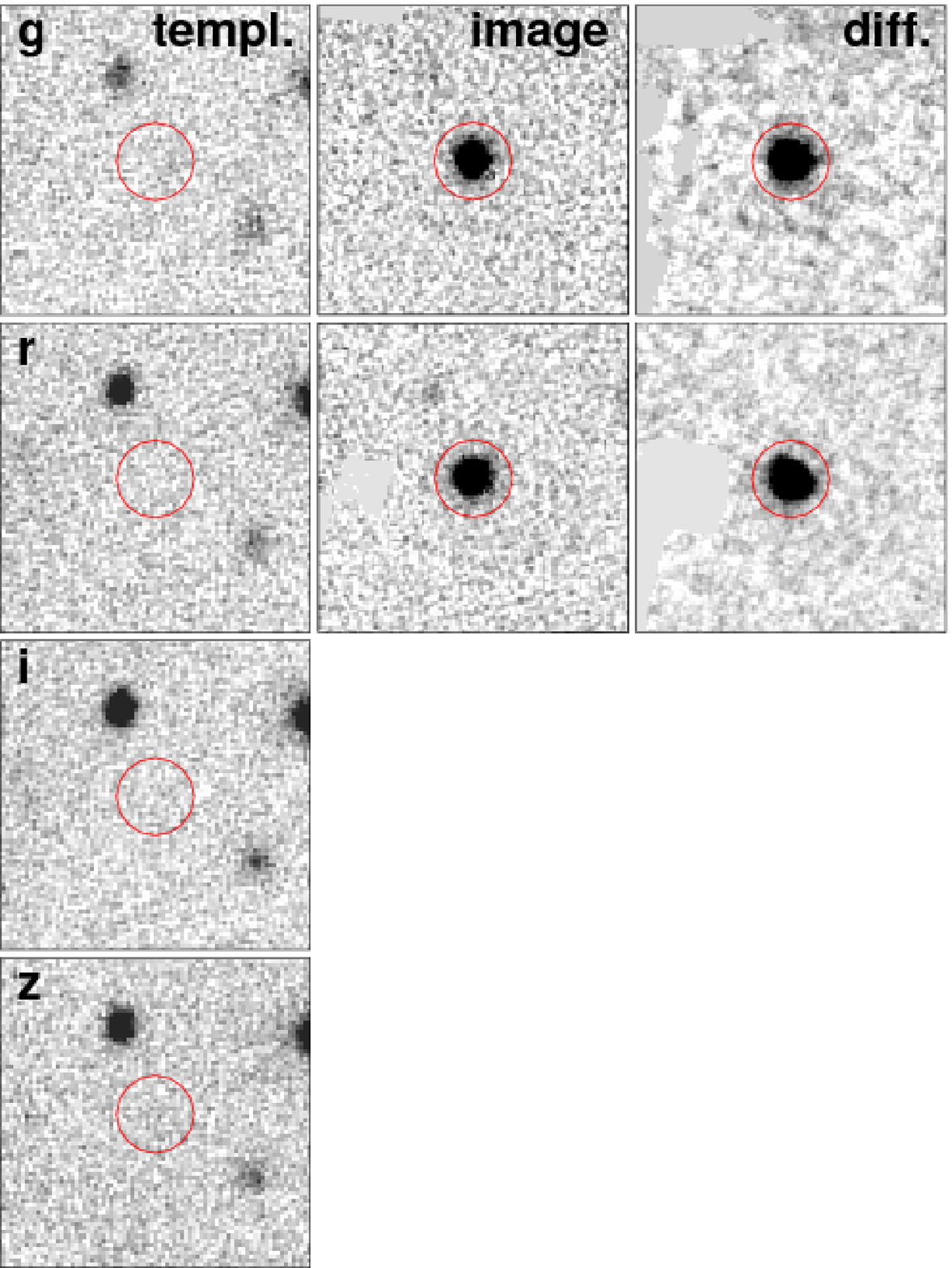}
\caption{PS1/MDS images of a representative M dwarf flare (left;
PSO\,J164.3814$+$58.3011) and an asteroid (right;
PSO\,J352.5968$-$0.4471) found in our fast transients search.  In each
case we show the template images (\gps\rps\ips\zps) in the left
column, the fast transient discovery images in the middle column, and
the difference images in the right column.  Each panel is $20''\times
20''$ oriented with north up and east to the left.  The red optical
colors of the M dwarf counterpart are evident, as is the lack of a
counterpart in the case of the asteroid.  The light curves resulting
from these detections are shown in Figure~\ref{fig:lcs}.
\label{fig:images}}
\end{figure*}

\section{Observations}
\label{sec:obs}

\subsection{PS1 Survey Summary}
\label{sec:ps1}

The PS1 telescope, located on Mount Haleakala, is a high-etendue
wide-field survey instrument with a 1.8-m diameter primary mirror and
a $3.3^\circ$ diameter field-of-view imaged by an array of sixty
$4800\times 4800$ pixel detectors, with a pixel scale of $0.258''$
\citep{PS1_system,PS1_GPCA}.  The observations are obtained through
five broad-band filters (\grizy), with some differences relative to
the Sloan Digital Sky Survey (SDSS); the \gps\ filter extends $200$
\AA\ redward of $g_{\rm SDSS}$ to achieve greater sensitivity and
lower systematics for photometric redshifts, while the \zps\ filter
terminates at $9300$ \AA, unlike $z_{\rm SDSS}$ which is defined by
the detector response \citep{tsl+12}.  PS1 photometry is in the
``natural'' system, $m=-2.5{\rm log}(F_\nu)+m'$, with a single
zero-point adjustment ($m'$) in each band to conform to the AB
magnitude scale.  Magnitudes are interpreted as being at the top of
the atmosphere, with 1.2 airmasses of atmospheric attenuation included
in the system response function \citep{tsl+12}.

The PS1 Medium-Deep Survey (MDS) consists of 10 fields (each with a
single PS1 imager footprint) observed on a nearly nightly basis by
cycling through the five filters in $3-4$ nights to a typical
$5\sigma$ depth of $\sim 23.3$ mag in \gps\rps\ips\zps, and $\sim
21.7$ mag in \yps.  The MDS images are processed through the Image
Processing Pipeline (IPP; \citealt{PS1_IPP}), which includes
flat-fielding (``de-trending''), a flux-conserving warping to a
sky-based image plane, masking and artifact removal, and object
detection and photometry.  For the fast transient search described
here we produced difference images from the stacked nightly images
using the {\tt photpipe} pipeline \citep{rsb+05} running on the
Odyssey computer cluster at Harvard University.

\subsection{A Search for Fast Optical Transients}
\label{sec:search}

\begin{figure}
\centering
\includegraphics[angle=0,width=0.48\textwidth]{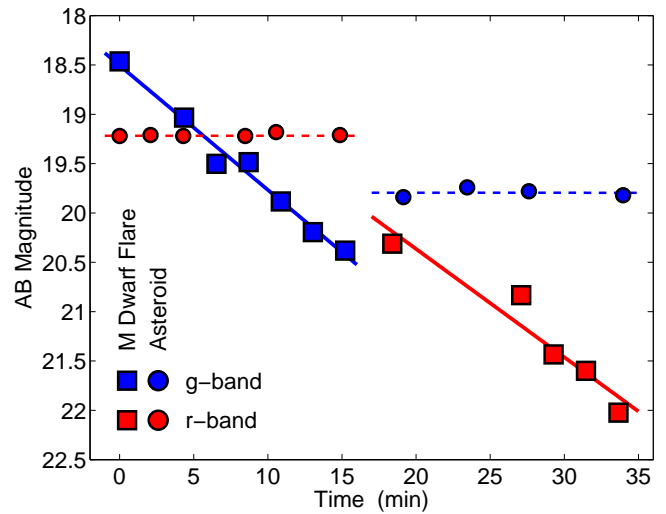}
\caption{Light curves in the \gps\ (blue) and \rps\ (red) filters for
a representative M dwarf flare (squares) and an asteroid (circles)
detected in our search for fast optical transients (see
Figure~\ref{fig:images}).  In each case the lines mark the best linear
fit.  The M dwarf flare exhibits a 3.5 mag decline during the
time-span of our observation, while the asteroid exhibits constant
brightness.  Missing light curve points are due to individual
exposures in which the source was located in a chip gap.
\label{fig:lcs}}
\end{figure}

For the purpose of detecting fast optical transients we take advantage
of the consecutive MDS \gps\rps\ observations, with eight $113$ s
exposures in each filter providing a total time-span of about $33$ min
for a full sequence.  We carry out the search using the stacked \gps\
and \rps\ images from each visit through image subtraction relative to
deep multi-epoch templates, and subsequently utilize the individual
exposures to construct light curves; representative discovery and
template images are shown in Figure~\ref{fig:images} and light curves
are shown in Figure~\ref{fig:lcs}.  We limit the timescale of the
transients to $\lesssim 1$ d by further requiring no additional
detections in the \gps\rps\ips\zps\ filters on preceding and
subsequent nights (extending to $\pm 5$ nights).  To ensure that this
constraint is met we only perform our search on the subset of MDS data
for which consecutive nights of observations are available.  In the
first $1.5$ years of data we searched a total of 277 nights of
\gps\rps\ observations across the 10 MDS fields, leading to a total
areal exposure of 40.4 deg$^2$ d for a timescale of 0.5 hr and 1940
deg$^2$ d for a timescale of 1 d.

To select transients in the \gps\rps\ difference images we utilize a
signal-to-noise ratio threshold\footnotemark\footnotetext{We
empirically correct the correlated noise in the difference image
detections by measuring the flux and uncertainty at random positions
in the difference images in the same manner as the transient flux, and
then determining a correction factor which leads to a distribution
with a reduced $\chi^2$ of unity.} of ${\rm S/N}=10$, with resulting
limiting magnitudes of $g_{\rm P1}\approx 22.7$ mag and $r_{\rm
P1}\approx 22.4$ mag.  We additionally require sources in the
\gps\rps\ difference images to astrometrically match within $0.35''$,
corresponding to a 1.2 pixel radius around each detection (this is
about 6 times the typical astrometric error; see
\S\ref{sec:asteroids}).  Using these cuts we find a total of 227
candidates, which were visually inspected by one of us (C.L.)  leading
to a final list of 19 sources that were further validated by two of us
(E.B.~and R.C.); the remaining 208 events were predominantly spurious
detections near saturated stars.

\section{The Properties of Pan-STARRS1 Fast Optical Transients}
\label{sec:results}

\begin{figure}
\centering
\includegraphics[angle=0,width=0.48\textwidth]{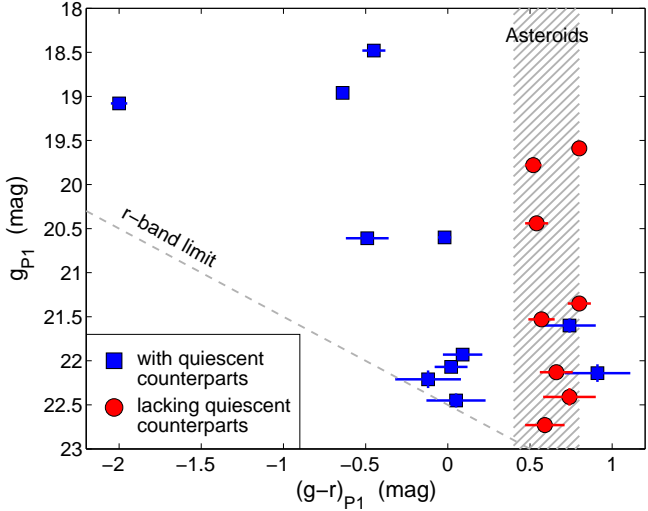}
\caption{Color-magnitude diagram for the fast transients discovered in
our PS1/MDS search.  The sources are divided into those with detected
quiescent counterparts in the deep template images (blue squares), and
those lacking counterparts to limits of \gps\rps\ips\zps$\gtrsim
24.5-25$ mag (red circles).  The dashed line marks the \rps\ limit of
our search (the region below the line is inaccessible to the survey).
The hatched region marks the expected $(g-r)_{\rm P1}$ color range for
asteroids.  All of the fast transients lacking quiescent counterparts
reside in this color range.
\label{fig:mag_all}}
\end{figure}

Using the procedure described in the previous section we found 19
genuine fast transients, spanning a brightness range of $g_{\rm P1}
\approx 18.5-22.7$ mag and $r_{\rm P1}\approx 18.8-22.4$ mag.  In
Figure~\ref{fig:mag_all} we show the \gps\ versus $(g-r)_{\rm P1}$
color-magnitude diagram for all 19 sources; the photometry is
summarized in Tables~\ref{tab:mdwarfs} and \ref{tab:asteroids}.  The
faint end of the distribution is determined by our requirement of
${\rm S/N}\gtrsim 10$ in the subtractions of the individual \gps\rps\
nightly stacks (16.5 min in each filter) from the deep templates.  On
the other hand, the bright end of the observed distribution is about 2
mag fainter than the saturation limit of our images, indicating a
genuine dearth of apparently bright fast transients in our search
area.  The $(g-r)_{\rm P1}$ colors span a wide range of about $-2.0$
to $+0.9$ mag, but we stress that for transients that rapidly vary in
brightness within the time-span of each observation (e.g.,
Figure~\ref{fig:lcs}), the non-simultaneous \gps\ and \rps\
measurements do not reflect the true instantaneous colors.

\subsection{Fast Transients Lacking Quiescent Counterparts}
\label{sec:asteroids}

\begin{figure}
\centering
\includegraphics[angle=0,width=0.48\textwidth]{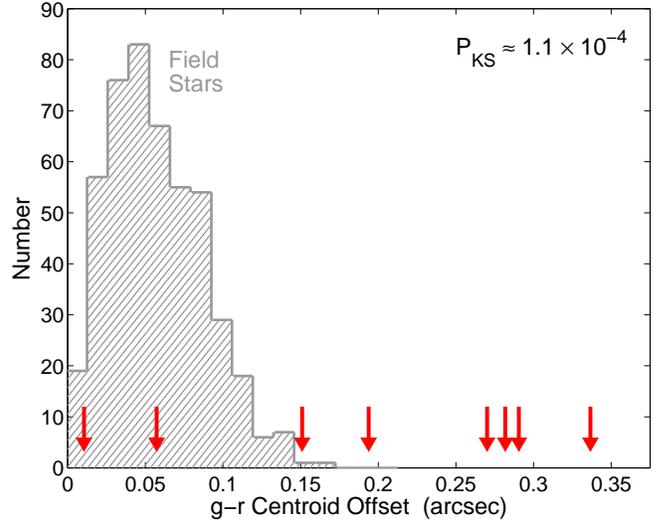}
\caption{Astrometric shift between the \gps\ and \rps\ centroids for
the fast transients lacking quiescent counterparts (red arrows) in
comparison to the shifts for field stars from the same images (gray
hatched histogram).  The median positional shift for the field stars,
which indicates the typical astrometric uncertainty of the MDS images,
is about 53 mas.  The fast transients lacking quiescent counterparts
exhibit generally larger shifts of up to $\approx 340$ mas (the limit
allowed by our search), with a median of 230 mas.  A K-S test
indicates a $p$-value of only $1.1\times 10^{-4}$ for the null
hypothesis that the two populations are drawn from the same underlying
distribution.
\label{fig:astrom}}
\end{figure}

Of the 19 fast transients discovered in our search, 8 events lack
quiescent counterparts in any of the deep template images
(\gps\rps\ips\zps) to typical limits of $\gtrsim 24.5-25$ mag.  These
sources are in principle a promising population of distant
extragalactic transients with undetected host galaxies.  We utilize a
combination of color information, the 33 min time-span of the
\gps\rps\ observations, and the sky locations to investigate the
nature of these sources.  We first note that all 8 sources span a
narrow color range with $(g-r)_{\rm P1}\approx 0.55-0.8$ mag
(Figure~\ref{fig:mag_all} and Table~\ref{tab:asteroids}), which is
typical of main-belt asteroids (e.g., \citealt{itr+01}).  This
interpretation naturally explains the lack of quiescent counterparts
in the template images.  We further test this scenario by comparing
the astrometric positions of each transient in the \gps\ and \rps\
images.  In Figure~\ref{fig:astrom} we show the distribution of
positional shifts for the 8 sources compared to unresolved field
sources in the same images.  The median offset for field sources is
about 53 mas (with a standard deviation of about 31 mas), indicative
of the astrometric alignment precision of the MDS images.  On the
other hand, the 8 transients exhibit shifts of $11-340$ mas (bounded
by our initial cut of $\lesssim 0.35''$ shift; \S\ref{sec:search}),
with a median value of about 230 mas.  This is well in excess of the
point source population, and a Kolmogorov-Smirnov (K-S) test gives a
$p$-value of only $1.1\times 10^{-4}$ for the null hypothesis that the
positional offsets of the 8 transients and the field sources are drawn
from the same underlying distribution.  This clearly indicates that
the 8 transients lacking quiescent counterparts exhibit larger than
average astrometric shifts, supporting their identification as
asteroids.  We further inspect the \gps\ and \rps\ light curves of the
8 transients and find that none exhibit variability larger than the
photometric uncertainties (e.g., Figure~\ref{fig:lcs}).

Finally, we note that all 8 sources are located in the three MDS
fields (MD04, MD09, and MD10) that are positioned within $\pm
10^\circ$ of the ecliptic plane (Table~\ref{tab:asteroids}).  In
particular, the asteroids were discovered in these fields on dates
concentrated at solar elongation values of about $130$ deg, at which
main-belt asteroids go through a stationary point with negligible
apparent motion.  We therefore conclude based on their colors,
astrometric motions, light curve behavior, and ecliptic coordinates
that the 8 fast transients lacking quiescent counterparts are simply
main-belt asteroids near the stationary point of their orbit.

\subsection{Fast Transients with Detected Quiescent Counterparts}
\label{sec:mdwarfs}

\begin{figure}
\centering
\includegraphics[angle=0,width=0.48\textwidth]{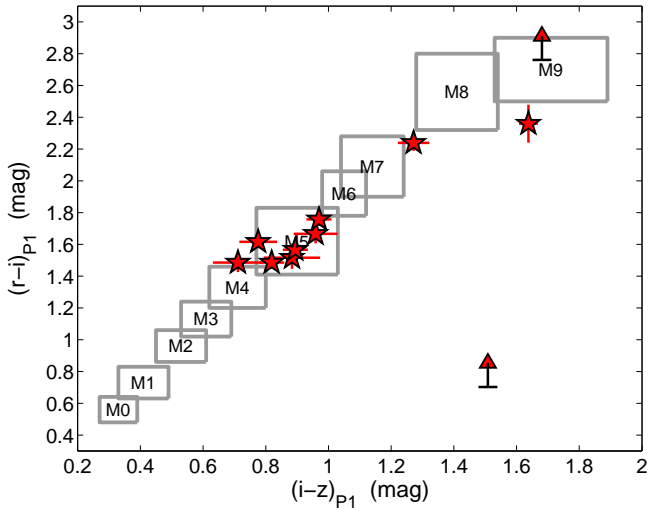}
\caption{Color-color phase-space for the quiescent counterparts of all
11 fast transients with a detected counterpart (stars; arrows indicate
limits).  Also shown are the color ranges for M0--M9 dwarf stars
\citep{wmb+11}.  All of the detected counterparts track the M dwarf
sequence, with 7 sources exhibiting the colors of M4--M5 dwarfs and 4
sources exhibiting colors typical of M7--M9 dwarfs.
\label{fig:mdwarfs}}
\end{figure}

We now turn to the 11 fast transients that exhibit quiescent
counterparts in some or all of the deep MDS template images.  In all
cases we find that the counterparts are unresolved (with a typical
seeing of about $1''$) and have red colors that are indicative of M
dwarf stars.  Photometry of the quiescent counterparts from the
PS1/MDS templates, and from SDSS when available, is summarized in
Table~\ref{tab:mdwarfs}.  Using these magnitudes we determine the
spectral type of each counterpart by comparing to the SDSS colors of M
dwarfs \citep{wmb+11}; from sources with both PS1 and SDSS photometry
we infer color transformations of $(g-r)_{\rm P1}\approx 0.94\times
(g-r)_{\rm SDSS}$ and $(i-z)_{\rm P1}\approx 0.93\times (i-z)_{\rm
SDSS}$ to account for the difference between the \gps and \zps\
filters compared to the $g_{\rm SDSS}$ and $z_{\rm SDSS}$ filters
(\S\ref{sec:ps1}).  The results are shown in Figure~\ref{fig:mdwarfs}
indicating that 7 counterparts have spectral types of about M4--M5,
while the remaining 4 counterparts have spectral types of M7--M9 (see
Table~\ref{tab:mdwarfs} for the inferred spectral types).  We further
infer the distances to these M dwarfs using their associated absolute
magnitudes \citep{bhw11} and find $d\approx 0.2-1.2$ kpc
(Table~\ref{tab:mdwarfs}).

\begin{figure}
\centering
\includegraphics[angle=0,width=0.48\textwidth]{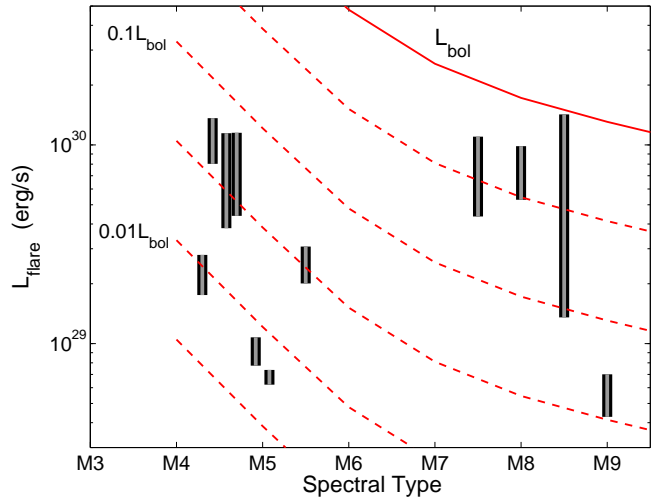}
\caption{Luminosities of the fast transients (flares) associated with
M dwarfs counterparts (vertical bars) as a function of source spectral
type.  The luminosity range for each source is defined by the \gps\
and \rps\ detections.  The solid red line marks the bolometric
luminosity as a function of spectral type, while dashed lines mark
fractions of the bolometric luminosity as indicated in the figure.  We
find that the flares from the $\sim{\rm M5}$ sources have a range of
$\approx 0.006-0.07$ L$_{\rm bol}$, while the flares from the M7--M9
dwarfs are relatively larger, with $\approx 0.03-1$ L$_{\rm bol}$.
\label{fig:flares}}
\end{figure}

Thus, as in previous fast transient searches, all 11 fast transients
with quiescent counterparts in our survey are M dwarf flares.  Using
the flare magnitudes and inferred spectral types, we compare the
resulting flare and bolometric luminosities in
Figure~\ref{fig:flares}.  We find that the flares span a
luminosity\footnotemark\footnotetext{We determine the luminosity by
integrating the spectral luminosity over the widths of the \gps\ and
\rps\ filters, with $\delta\nu\approx 1.287\times 10^{14}$ Hz and
$\approx 7.721\times 10^{13}$ Hz, respectively.}  range of
$L_{f,g}\approx (6-150)\times 10^{28}$ erg s$^{-1}$ and
$L_{f,r}\approx (4-80)\times 10^{28}$ erg s$^{-1}$, with no apparent
dependence on spectral type.  However, since the bolometric luminosity
declines from about $3.3\times 10^{31}$ erg s$^{-1}$ at spectral type
M4 to about $1.3\times 10^{30}$ erg s$^{-1}$ at spectral type M9, the
relative flare luminosities increase with later spectral type.  We
find relative flare luminosities of $\approx 0.006-0.07$ L$_{\rm bol}$
at $\sim{\rm M5}$, and larger values of $\approx 0.03-1$ L$_{\rm bol}$
at M8--M9.

It is instructive to compare the properties of the flares and M dwarfs
uncovered in our blind fast transients search to those from targeted M
dwarf variability studies.  In particular, \citet{khh+09} searched for
flares from 50,130 pre-selected M0--M6 dwarfs in the SDSS Stripe 82
and found 271 flares, with an apparent increase in the flare rate with
later spectral type.  This may explain the lack of M0--M4 dwarfs in
our relatively small sample.  For the M4--M6 dwarfs \citet{khh+09}
find a mean flare amplitude of $\Delta u\approx 1.5$ mag, comparable
to our mean value of $\Delta g\approx 1.1$ mag, when taking into
account that M dwarfs flares are generally brighter in $u$-band than
in $g$-band due a typical temperature of $\sim 10^4$ K.  The flare
luminosities for the Stripe 82 M4--M6 dwarfs are $L_{f,u}\approx
(2-100)\times 10^{28}$ erg s$^{-1}$, again comparable to the \gps-band
luminosities of the $\sim {\rm M5}$ dwarfs in our sample.  We note
that there are no M7--M9 dwarfs in the Stripe 82 sample.

\citet{khh+09} also found that for M4--M6 dwarfs there is a strong
dependence of the flare rate on vertical distance from the Galactic
plane; namely, the fraction of time in which a star flares decreases
by about an order of magnitude over a vertical distance range of about
$50-150$ pc.  The M4--M6 dwarfs in our sample are all located at
larger vertical distances of $\approx 190-560$ pc, with a mean of
about 390 pc, suggesting that the decline in flaring activity with
vertical distance from the Galactic plane may not be as steep as
previously inferred.  Moreover, the Stripe 82 data exhibit a trend of
steeper decline in the flare rate as a function of vertical distance
with increasing spectral type \citep{khh+09}, while here we find that
the M7--M9 dwarfs have a similar mean vertical distance from the
Galactic plane to the $\sim{\rm M5}$ dwarfs.  Clearly, a more
systematic search for M dwarf flares in the PS1/MDS is required to
study these trends.  In particular, it is likely that we have missed
some flares due to the requirement of no additional variability within
a $\pm 5$ night window around each detection (\S\ref{sec:search}).

\section{Limits on the Rate of Extragalactic Fast Optical Transients}
\label{sec:rate}

The 19 fast transients uncovered by our search cleanly divide into two
categories: (i) main-belt asteroids near the stationary point of their
orbits (\S\ref{sec:asteroids}); and (ii) flares from M dwarf stars
(\S\ref{sec:mdwarfs}).  Neither category is unexpected given that our
search is based on consecutive \gps\rps\ detections with a time-span
of about 0.5 hr.  M dwarf flares typically exhibit blue colors
indicative of $T\sim 10^4$ K, with timescales of minutes to hours, and
are thus ubiquitous in searches that utilize rapid observations in the
ultraviolet \citep{wwh+05,wwb+06} or blue optical bands
\citep{bwb+04,kr06,rok+08}.  Similarly, our requirement of two
consecutive detections within $\sim 0.5$ hr, with non-detections on
preceding or subsequent nights is effective at capturing asteroids
near the stationary point of their orbits (i.e., at solar elongations
of about $130$ deg for main-belt asteroids).

\begin{figure*}
\centering
\includegraphics[angle=0,width=0.92\textwidth]{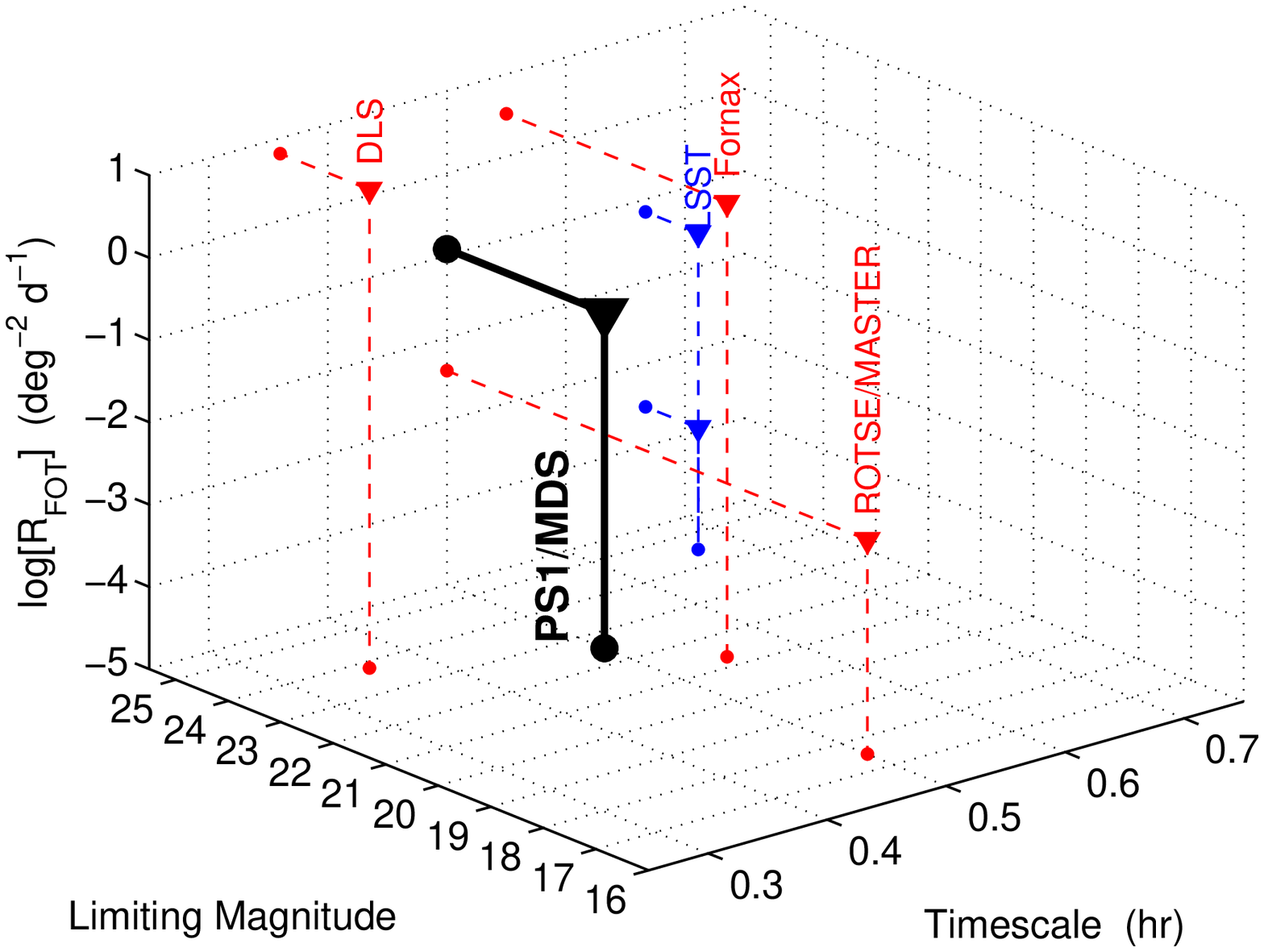}
\caption{Limits on the sky-projected rate of extragalactic fast
optical transients as a function of timescale and survey limiting
magnitude.  Shown are the limits from our survey (black) and from the
literature (red; \citealt{bwb+04,raa+05,lkk+07,rok+08}).  The much
larger effective areal exposure of our survey provides constraints
that are about $30-50$ times deeper than previous surveys with similar
limiting magnitudes.  Also shown are the expected limits from the LSST
main survey (blue) for 1 night (top limit) and 1 year (bottom limit)
of observations.
\label{fig:surveys1}}
\end{figure*}

Since we account for all 19 fast transients as Solar system or
Galactic in origin, we can place a robust upper limit on the rate of
extragalactic fast optical transients.  While we could in principle
detect a sufficiently bright transient with a timescale as short as
about 4 min, corresponding to a detection in only the final exposure
in one filter and the first exposure in the second filter, a more
reasonable timescale probed by our search is about 0.5 hr, the
time-span of a full \gps\rps\ exposure sequence (e.g.,
Figure~\ref{fig:lcs}).  Similarly, our selection criteria could in
principle accommodate transients with durations as long as $\sim 2$ d,
but for the bulk of the search the maximum timescale is $\lesssim 1$
d.  Thus, we consider our search to place limits on fast transients
spanning about 0.5 hr to 1 d.  In the discussion below we provide
upper limits for the upper and lower bounds of the timescale
distribution.

The total areal exposure of our survey for a timescale of 0.5 hr is
$E_A\approx 40.4$ deg$^{2}$ d, while for a timescale of $\sim {\rm
day}$ it is correspondingly longer, $E_A\approx 1940$ deg$^{2}$ d.
Thus, we can place a $95\%$ confidence limit ($\lesssim 3$ events) on
the sky-projected rate of:
\begin{equation} 
R_{\rm FOT}=N/\epsilon^2E_A, 
\end{equation} 
where we estimate the detection efficiency per filter at $\epsilon
\approx 0.8$ based on the overall search for transients in the MDS
fields.  Thus, for a timescale of 0.5 hr we place a limit of $R_{\rm
FOT}\lesssim 0.12$ deg$^{-2}$ d$^{-1}$, while for a timescale of $\sim
{\rm day}$ it is $R_{\rm FOT}\lesssim 2.4\times 10^{-3}$ deg$^{-2}$
d$^{-1}$ (see Table~\ref{tab:rates}).

Our limit on fast transients with a 0.5 hr timescale improves on
existing searches with similar limiting magnitudes by a factor of
$30-50$ (Figure~\ref{fig:surveys1}) thanks to the much larger
effective areal exposure.  DLS ($E_A\approx 1.1$ deg d) placed an
upper limit of $R_{\rm FOT}\lesssim 6.5$ events deg$^{-2}$ d$^{-1}$
for transients with a timescale of about $0.36$ hr \citep{bwb+04},
while the Fornax galaxy cluster search ($E_A\approx 1.9$ deg d) placed
a limit of $R_{\rm FOT}\lesssim 3.3$ deg$^{-2}$ d$^{-1}$ for
transients with a timescale of about $0.55$ hr \citep{rok+08}.  The
ROTSE-III search ($E_A\approx 635$ deg d) placed a limit of $R_{\rm
FOT}\lesssim 5\times 10^{-3}$ deg$^{-2}$ d$^{-1}$ for transients with
a timescale of about $0.5$ hr, but was significantly shallower
\citep{raa+05}.

The sky-projected rates do not take into account the difference in
limiting magnitudes between various searches.  Our survey limiting
magnitude is about 1.3 mag shallower than the DLS search, but about
1.2 mag deeper than the Fornax cluster search, and about 5 mag deeper
than the ROTSE-III search.  This information is summarized in
Table~\ref{tab:rates}, and shown in the relevant three-dimensional
phase-space of survey limiting magnitude, timescale, and sky-projected
rate limit in Figure~\ref{fig:surveys1}.

The survey depth impacts the inferred limits on {\it volumetric}
rates.  For example, for a population of fast transients with a
fiducial absolute magnitude of $-10$ (comparable to the most luminous
known novae; see \S\ref{sec:models}), the limits on the volumetric
rates\footnotemark\footnotetext{We assume a uniform underlying
distribution of galaxies within the volume probed by each search,
which is not strictly the case for the maximum distances associated
with a peak magnitude of $-10$: 32 Mpc for our search, 56 Mpc for DLS,
and 18 Mpc for Fornax.  With the exception of the Fornax search, which
was centered on the Fornax galaxy cluster at $d\approx 16$ Mpc, this
indicates that the actual limits from our survey and from DLS are
subject to the underlying non-uniform galaxy distribution.} on a
timescale of 0.5 hr are $\lesssim 13$ Mpc$^{-3}$ yr$^{-1}$ for our
survey, $\lesssim 1.3\times 10^2$ Mpc$^{-3}$ yr$^{-1}$ for DLS, and
$\lesssim 2.0\times 10^3$ Mpc$^{-3}$ yr$^{-1}$ for Fornax.  The small
volume probed by the ROTSE-III search for transients with $-10$ mag
($d\approx 3$ Mpc) provides no real insight on the extragalactic
population.  For a fiducial absolute magnitude of $-14$ (comparable to
the least luminous supernovae), the limits are $\lesssim 0.05$
(PS1/MDS), $\lesssim 0.5$ (DLS), and $\lesssim 7.8$ (Fornax)
Mpc$^{-3}$ yr$^{-1}$; at this peak magnitude the maximal detection
distances are large enough to uniformly sample the galaxy distribution
(about $200$ Mpc for our search; $360$ Mpc for DLS, and $115$ Mpc for
Fornax).  Finally, for a fiducial absolute magnitude of $-24$
(intermediate between long and short gamma-ray burst afterglows; see
\S\ref{sec:models}), the volumetric rate limits are $\lesssim 10^{-6}$
(PS1/MDS), $\lesssim 6\times 10^{-5}$ (DLS), and $\lesssim 3\times
10^{-5}$ (Fornax) Mpc$^{-3}$ yr$^{-1}$.  The various limits are
summarized in Table~\ref{tab:rates}.

We note that the inferred limits at $-10$ and $-14$ mag are
substantially higher than the volumetric rate of supernovae, $\approx
10^{-4}$ Mpc$^{-3}$ yr$^{-1}$ \citep{lcl+11}, and
novae\footnotemark\footnotetext{We use a nova rate of $2.2$ yr$^{-1}$
per $10^{10}$ L$_{\rm K,\odot}$ \citep{ws04}, combined with the
$K$-band luminosity density in the local universe of $4.4\times 10^8$
L$_{\rm K,\odot}$ Mpc$^{-3}$ \citep{cnb+01,bmk+03}.}, $\approx 0.1$
Mpc$^{-3}$ yr$^{-1}$; similarly, the limits at $-24$ mag are at least
two orders of magnitude larger than the on-axis GRB rate.  This
indicates that any source population of extragalactic fast optical
transients with a timescale of 0.5 hr would have to be much more
abundant than nova, supernova, or GRB progenitors, or produce multiple
($\sim 10^2-10^3$) events per progenitor system to be detected with
current surveys.

Utilizing the $B$-band luminosity density in the local universe
($\approx 1.4\times 10^8$ L$_{\rm B,\odot}$ Mpc$^{-3}$) the upper
limits on the volumetric rates at $-10$ mag can be recast as $\lesssim
9.2\times 10^2$ (PS1/MDS) and $\lesssim 9.3\times 10^3$ (DLS)
yr$^{-1}$ per $10^{10}$ L$_{\rm B,\odot}$; for the Fornax survey,
which targeted a galaxy cluster environment, the limit is $\lesssim
350$ yr$^{-1}$ per $10^{10}$ L$_{\rm B,\odot}$ \citep{rok+08}.  For a
fiducial brightness of $-14$ mag, the limits are $\lesssim 3.6$
(PS1/MDS), $\lesssim 36$ (DLS), and $\lesssim 75$ yr$^{-1}$ per
$10^{10}$ L$_{\rm B,\odot}$; the latter value is from \citet{rok+08}.

For the fiducial timescale of about 1 d (the upper bound of our
survey), the volumetric rate limits are $\lesssim 0.3$ ($-10$ mag),
$\lesssim 10^{-3}$ ($-14$ mag), and $\lesssim 4\times 10^{-8}$ ($-24$
mag) Mpc$^{-3}$ yr$^{-1}$.  The rates at $-14$ and $-24$ mag are of
interest since they are only an order of magnitude larger than the
supernova and GRB rates, indicating that an expansion of our search
might yield interesting limits on the rate of fast transients with
$\sim 1$ d timescale.

\section{Theoretical Expectations for the Luminosity of Fast Optical
Transients} 
\label{sec:models}

In the discussion above we used fiducial peak absolute magnitudes of
$-10$, $-14$, and $-24$ mag to infer limits on the volumetric rates of
extragalactic fast optical transients.  These particular values were
chosen to match the most luminous classical novae, the least luminous
supernovae, and typical long and short GRB afterglows, respectively.
However, it is instructive to explore what peak luminosities are
expected for sources with a fiducial timescale of $\sim 0.5$ hr to
$\sim{\rm day}$ for a range of potential mechanisms.  We defer an
exhaustive investigation of this question to future work and focus on
the basic arguments here.  At the most basic level, we note that
explosive sources with a characteristic velocity of $\sim 10^4$ km
s$^{-1}$ that emit thermally with a peak in the optical ($T\sim 10^4$
K) will be limited to an absolute magnitude of $\gtrsim -14$; only
sources undergoing relativistic expansion (e.g., GRBs) may
significantly exceed this limits.  We expand on this general theme
below.

We first investigate the possibility of unusually fast and luminous
classical novae.  In general, novae follow the maximum magnitude
versus rate of decline (MMRD) relation (e.g., \citealt{dl95}), which
indicates that faster novae are also more luminous.  However, the MMRD
relation flattens to a maximal observed value of about $-10$ mag for
timescales shorter than a few days \citep{dl95}.  Indeed, the shortest
observed timescales (quantified as $t_2$ or $t_3$, the timescales to
decline by 2 or 3 magnitudes, respectively) are $\approx 3-5$ d
\citep{cbc+13}.  Theoretical models point to maximal magnitudes of
about $-10$ and timescales as short as $t_2\approx 1$ d
\citep{yps+05}.  Moreover, the fraction of novae that achieve such
high peak luminosity and rapid fading is $\lesssim 1\%$ (e.g.,
\citealt{srq+09}), leading to a rate of $\lesssim 10^{-3}$ Mpc$^{-3}$
yr$^{-1}$, orders of magnitude below the upper bounds for $-10$ mag
derived from the various searches (e.g., 0.3 Mpc$^{-3}$ yr$^{-1}$ for
our search; Table~\ref{tab:rates}).  Thus, novae are not expected to
contribute a population of luminous and fast extragalactic transients
for current surveys.

\begin{figure*}
\centering
\includegraphics[angle=0,width=0.92\textwidth]{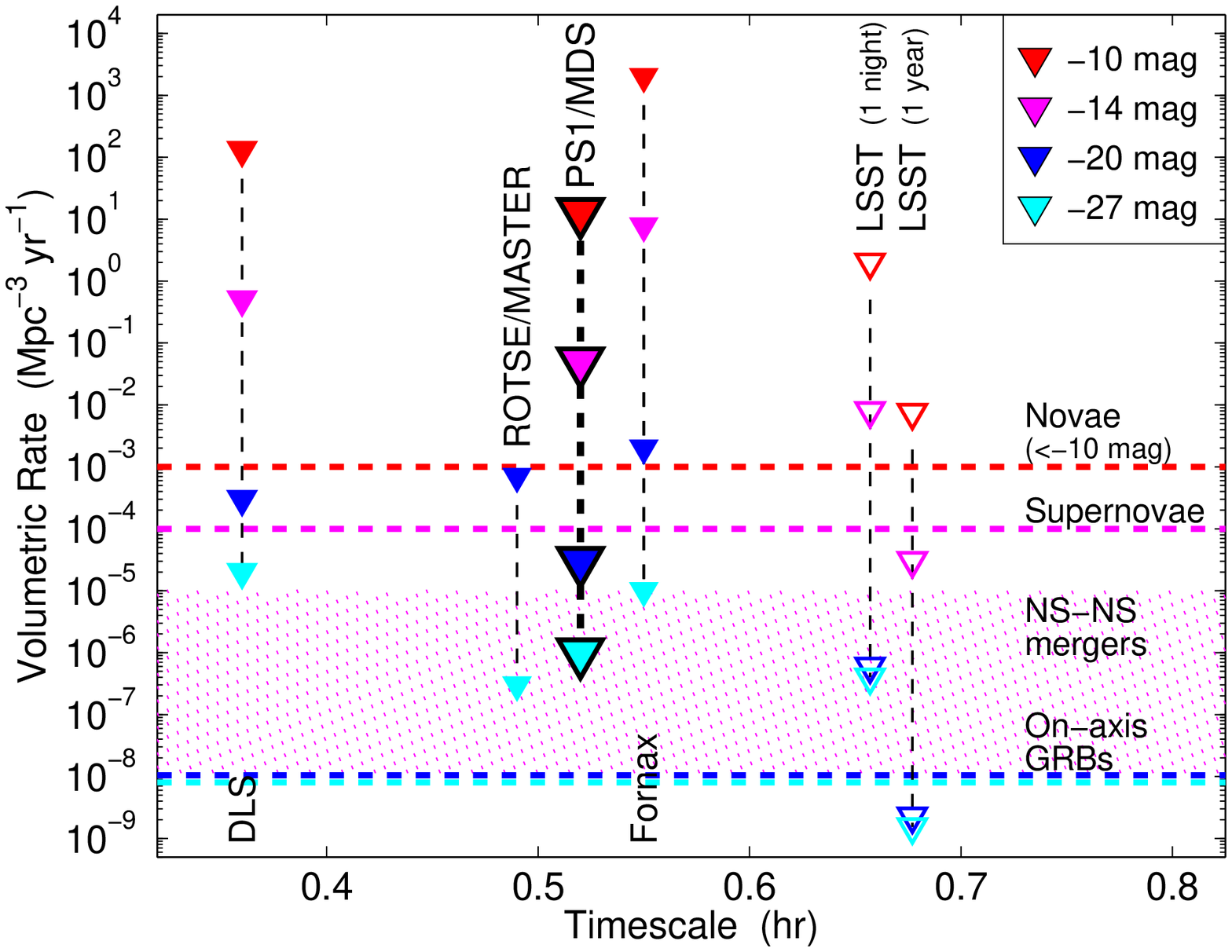}
\caption{Limits on the volumetric rate of extragalactic fast optical
transients as a function of timescale from our PS1/MDS survey and from
the literature \citep{bwb+04,raa+05,lkk+07,rok+08}.  For each survey
we provide limits for a range of fiducial absolute magnitudes (top to
bottom) of $-10$ (most luminous novae), $-14$ (least luminous SNe),
$-20$ (on-axis short GRB), and $-27$ (on-axis long GRB).  Also marked
are the actual volumetric rates of luminous novae, supernovae, binary
neutron star mergers, and on-axis GRBs.  The inferred limits from all
existing surveys are orders of magnitude larger than the known
volumetric rates, indicating that large new populations of
currently-unknown astrophysical fast transients are required for
detectability, or that much larger surveys are essential to produce
meaningful limits.  The single night and full year limits for the LSST
main survey indicate that this survey should detect some fast optical
transients (the LSST estimates have been shifted along the time axis
for clarity).
\label{fig:surveys2}}
\end{figure*}

We next explore the generic case of an explosive event with ejecta
that cool adiabatically due to free expansion ($R=v_{\rm ej}t$); i.e.,
a supernova-like explosion, but lacking internal heating.  In this
case, with the expansion driven by radiation pressure and using the
standard diffusion approximation \citep{arn82}, the luminosity is
given by:
\begin{equation}
L(t)=L_0\,{\rm exp}\left[-\frac{t}{\tau_{d,0}}-\frac{t^2}{\sqrt{2}
\tau_{d,0} \tau_h}\right],
\end{equation}
where $\tau_h=R_0/v_{\rm ej}$ is the initial hydrodynamic timescale,
$\tau_{d,0}=B\kappa M_{\rm ej}/cR_0$ is the initial diffusion
timescale, $\kappa$ is the opacity (for simplicity we use
$\kappa=\kappa_{\rm es}=0.4$ cm$^2$ g$^{-1}$), $M_{\rm ej}$ is the
ejecta mass, $B\approx 0.07$ is a geometric factor, and
$L_0=E_0/\tau_{d,0}$ is the initial luminosity.  A fiducial timescale
of 0.5 hr requires $R_0\lesssim 0.5\,{\rm hr}\times v_{\rm ej}\approx
2\times 10^{12}\,(v_{\rm ej}/10^4\,{\rm km\,s^{-1}})$ cm; i.e., more
extended sources are unlikely to produce transients with a duration as
short as $\sim 0.5$ hr.  Using the appropriate size limit we find that
for a source with $L(0.5\,{\rm hr})/L_0\sim 0.1$ the resulting peak
bolometric magnitude ranges from about -8 mag (for a white dwarf with
$R_0\approx 10^9$ cm) to about $-15$ mag (for a star with $R_0\approx
10$ R$_\odot$).  The resulting characteristic effective temperature
ranges from $T_{\rm eff}=(L_0/4\pi R^2 \sigma)^{1/4}\approx 2.2\times
10^4$ K (for $R_0\approx 10^9$ cm) to $\approx 1.1\times 10^5$ K (for
$R_0\approx 10$ R$_\odot$), or a peak in the ultraviolet.  This means
that the peak absolute magnitude in $gr$ will be a few magnitudes
fainter, corresponding to the faint end of the magnitude range
considered in the previous section (i.e., $\gtrsim -14$ mag).

The luminosity can be enhanced by appealing to an internal energy
source, for example radioactive heating as in the case of Type I
supernovae.  In this scenario the luminosity is maximized if the
radioactive decay timescale is well-matched to the diffusion
timescale, $\tau_r\approx\tau_d\approx B\kappa M_{\rm ej}/cR\approx
0.5$ hr.  The fast timescale investigated here requires different
radioactive material than $^{56}$Ni, which powers the optical light
curves of Type I supernovae, since the latter has $\tau_r\approx 8.8$
d.  Instead, a better match may be provided by the radioactive decay
of $r$-process elements, for which there is a broad range of
timescales (e.g., \citealt{lp98,mmd+10}).  Using a timescale of 0.5 hr
we find that the required ejecta mass is low, $M_{\rm ej}\approx
2\times 10^{-6}\,(v_{\rm ej}/10^4\,{\rm km/s})^2$ M$_\odot$.  For
radioactive heating the total energy generation rate is
$\epsilon(t)=(fM_{\rm ej} c^2/\tau_r){\rm e}^{-t/\tau_r}$, where $f\ll
1$ is an efficiency factor, likely in the range of $10^{-6}-10^{-5}$
\citep{mmd+10}.  For the ejecta mass inferred above we find
$\epsilon(t)\approx 10^{45}\,f$ erg s$^{-1}$, and hence $L\sim
10^{40}$ erg s$^{-1}$ (for $f\sim 10^{-5}$), or an absolute magnitude
of about $-12$ mag.  As in the discussion above, the spectrum will
peak in the ultraviolet, and the optical emission ($g$-band) will be
dimmer by about 2 mag (i.e., to a peak of about $-10$ mag).  For a
fiducial timescale of $\sim{\rm day}$, the peak absolute magnitude is
larger, reaching $\approx -14$ mag in the optical.  This is
essentially the ``kilonova'' model invoked for compact object binary
mergers \citep{lp98,mmd+10,bfc13}, with a predicted volumetric rate of
$\sim 10^{-8}-10^{-5}$ Mpc$^{-3}$ yr$^{-1}$ (e.g., \citealt{aaa+10}).
We note, however, that recent opacity calculations for $r$-process
material indicate that the actual peak brightness in the optical bands
will be an order of magnitude fainter than suggested by the above
calculation \citep{bk13}.  A similar scenario is the thermonuclear .Ia
supernova model with heating by radioactive $\alpha$-chain elements
(Ca, Ti), leading to a peak optical brightness of $\sim -15$ mag, but
with a longer timescale of $\sim{\rm week}$ even for low helium shell
masses \citep{skw+10}.  Thus, models with radioactive heating are
limited to $\sim -10$ mag ($\sim 0.5$ hr) or $\sim -14$ mag ($\sim{\rm
day}$).

An alternative internal energy source is the spin-down of a newly-born
millisecond magnetar \citep{kb10,woo10}.  In this scenario the maximal
luminosity is achieved when the magnetar spin-down timescale,
$\tau_{\rm sd}\approx 4\times 10^3\,(B/10^{14}\,{\rm G})^{-2}\,
(P/1\,{\rm ms})^2$ s, is comparable to the diffusion timescale,
$\tau_d\approx 0.5$ hr; here $B$ is the magnetic field strength and
$P$ is the initial rotation period.  The available rotational energy
is large, $E_{\rm rot}\approx 2\times 10^{52}\,(P/1\,{\rm ms})^{-2}$
erg, and therefore a magnetar engine can in principle produce
extremely luminous fast transients, with a peak brightness of
$L\approx E_{\rm rot}\,t_{\rm sd}/t_d^2\approx 5\times 10^{48}$ erg
s$^{-1}$.  However, since $\tau_d\approx 0.5$ hr requires a low ejecta
mass of $\sim 10^{-6}$ M$_\odot$, this scenario necessitates magnetar
birth with negligible ejected mass, and results in a relativistic
outflow; this is essentially the scenario invoked in magnetar models
of gamma-ray bursts.  It is not obvious in this scenario whether the
large energy reservoir will be emitted in the optical band, or
primarily at X-ray/$\gamma$-ray energies.  Thus, the optical signature
may still be rather weak, and perhaps dominated instead by interaction
with the ambient medium (i.e., an afterglow).  Overall, the estimated
magnetar birth rate is $\sim 10\%$ of the core-collapse supernova rate
\citep{kds+98}, while the GRB rate is $\lesssim 1\%$ of the supernova
rate \citep{wp10}.  This indicates that any magnetar-powered fast
optical transients will have a rate orders of magnitude below the
upper limits inferred by the existing searches.

Finally, luminous rapid optical transients can be produced by
circumstellar interaction of a relativistic outflow, such as an
on-axis\footnotemark\footnotetext{For significant off-axis angles the
optical afterglow emission is expected to increase for days and then
decline as a $\sim t^{-1}$ power law thereafter, leading to a
characteristic timescale much longer than 0.5 hr.}  GRB afterglow.
The typical absolute magnitude of a short-duration GRB afterglow at
$\delta t\approx 0.5$ hr is about $-20$ mag (based on a typical
apparent brightness of about $22$ mag and a typical redshift of
$z\approx 0.5$; \citealt{ber10}).  At this fiducial luminosity the
inferred volumetric upper limits from the fast optical transient
searches are $\lesssim 3\times 10^{-5}$ (PS1/MDS), $\lesssim 3\times
10^{-4}$ (DLS), $\lesssim 2\times 10^{-3}$ (Fornax), and $\lesssim
7\times 10^{-4}$ (ROTSE-III) Mpc$^{-3}$ yr$^{-1}$.  The actual
volumetric rate of on-axis short GRBs is about $10^{-8}$ Mpc$^{-3}$
yr$^{-1}$ \citep{ngf06}, at least $3\times 10^3$ times lower than the
limits reached by the surveys.  For long-duration GRBs the typical
absolute magnitude at $\delta t\approx 0.5$ hr is much larger, about
$-27$ mag (based on a typical apparent brightness of about $17$ mag
and a typical redshift of $z\approx 2$; \citealt{kkz+10}).  At this
luminosity the volumetric upper limits for the various searches
are\footnotemark\footnotetext{We note that the volume accessible to
our survey, as well as the DLS and the Fornax search is actually
limited by their use of the $B$-band filter, which provides
sensitivity only to $z\approx 4$ due to strong suppression by the
Lyman alpha forest.} $\lesssim 10^{-6}$ (PS1/MDS), $\lesssim 2\times
10^{-5}$ (DLS), $\lesssim 1\times 10^{-5}$ (Fornax), and $\lesssim
3\times 10^{-7}$ (ROTSE-III) Mpc$^{-3}$ yr$^{-1}$.  The inferred
on-axis volumetric rate (at $z\sim 2$) is $\sim 10^{-8}$ Mpc$^{-3}$
yr$^{-1}$ \citep{wp10}.  Thus, unless there is a substantial
population of relativistic explosions that do not produce $\gamma$-ray
emission, the existing limits are much too shallow in the context of
on-axis GRB rates.

We therefore conclude based on the comparison to various potential
models and populations that: 
\begin{itemize} 
\item {\it non-relativistic} fast optical transients with a timescale
of $\sim 0.5$ hr ($\sim{\rm day}$) are generically limited to a peak
optical brightness of $\sim -10$ ($\sim -14$) mag even if they are
powered by radioactive heating; revised opacities for $r$-process
matter may reduce these values by about an order of magnitude.
Similarly, unusually fast classical novae are not likely to reach peak
magnitudes larger than about $-10$ mag.  Overall, the limits achieved
by existing searches at these luminosities (Table~\ref{tab:rates}) are
orders of magnitude higher than the known and expected event rates for
such transients (e.g., $\lesssim 10^{-3}$ Mpc$^{-3}$ yr$^{-1}$ for
luminous novae).
\item {\it relativistic} sources such as on-axis GRBs or transients
powered by the spin-down of millisecond magnetars can produce much
larger luminosities on a timescale of $\sim 0.5$ hr, but the
volumetric rate limits from the existing searches are still orders of
magnitude larger than the very low known or anticipated rates of
relativistic explosions.  A possible exception is a fast transient
recently discovered by \citet{ckh+13}, although with a timescale of
$\sim {\it few}$ d.
\end{itemize}

In Figure~\ref{fig:surveys2} we summarize this information by
comparing the volumetric rate limits from the various surveys at
fiducial peak absolute magnitudes of $-10$, $-14$, $-20$ (short GRB),
and $-27$ (long GRB) to the rates of known transients (e.g., novae,
supernovae, neutron star binary mergers, on-axis short and long GRBs).
The results demonstrate that the existing survey limits are orders of
magnitude above the anticipated event rates based on the known
classes.  This indicates that much larger surveys are essential to
robustly explore the transient sky on the $\sim {\rm hour-day}$
timescale.  In addition, we stress that given the low expected
luminosity for non-relativistic fast transients ($\gtrsim -14$), a
profitable strategy is to systematically target nearby galaxies (as
was done for example in the Fornax Cluster search; \citealt{rok+08})
rather than image wide blank fields.  On the other hand, searches for
the rare but highly luminous relativistic fast transients are best
focused on covering wide fields (preferably $\sim{\rm all-sky}$) at
the expense of survey depth since the projected rate is at most $\sim
{\rm few}$ per sky per day, while the limiting factor for distance
coverage is the use of optical filters (limited to $z\sim 6$) rather
than depth.

\section{Implications for LSST} 
\label{sec:lsst}

Our search for fast optical transients proved highly effective at
identifying foreground events {\it without the need for expensive
follow-up observations}.  This is unlike previous searches such as DLS
and Fornax.  This has been accomplished thanks to the use of
dual-filter observations spanning a timescale of about 0.5 hr, with
each observation composed of multiple exposures that help to eliminate
individual spurious detections.  In addition, the availability of
color information ($g-r$ in this case), coupled with the time baseline
allowed us to effectively identify main-belt asteroids near the
stationary point.  Similarly, deep multi-band templates allowed us to
unambiguously identify M dwarf counterparts for all detected flares to
a distance of at least $\sim 1.2$ kpc and with spectral types
extending to about M9.  With this information we were able to account
for all the fast optical transients found in the survey, and to place
the deepest limit to date on the rate of extragalactic fast
transients.  Our results are reassuring given that follow-up
spectroscopy in the era of on-going and future large surveys is in
limited supply.

With this in mind, it is instructive to consider our results in the
context of the anticipated LSST survey strategy \citep{ita+08}.  The
LSST main survey is intended to cover a total of $\approx 18,000$
deg$^2$, accounting for about $90\%$ of the observing time.  Each
pointing will consist of two visits per field with a separation of
about $15-60$ min, with each visit consisting of a pair of 15 s
exposures with a $3\sigma$ depth of $\sim 24.5$ mag (or about 23.5 mag
at $10\sigma$).  It is anticipated that a total of about $3000$
deg$^2$ will be imaged on any given night, potentially with more than
one filter, leading to an areal exposure of about $62$ deg$^2$ d in a
single night, or about\footnotemark\footnotetext{Assuming $90\%$ of
the total observing time and a $20\%$ loss due to weather and
maintenance.} $1.6\times 10^4$ deg$^2$ d per year, for the fiducial
timescale of 0.5 hr.  This survey will therefore achieve a sky
projected rate limit ($95\%$ confidence level) of $0.07$ deg$^{-2}$
d$^{-1}$ in a single night and $3\times 10^{-4}$ deg$^{-2}$ d$^{-1}$
in a year.  The latter value represents a factor of 400 times
improvement relative to our PS1/MDS search.  In terms of a volumetric
rate, the greater depth achieved by LSST will probe rates of $\approx
9\times 10^{-3}$ Mpc$^{-3}$ yr$^{-1}$ for $-10$ mag and $\approx
3\times 10^{-5}$ Mpc$^{-3}$ yr$^{-1}$ for $-14$ mag for 1 year of
operations.  At a typical brightness of short and long GRBs, the rate
will be about $2\times 10^{-9}$ Mpc$^{-3}$ yr$^{-1}$ (limited to
$z\lesssim 6$ by the $zy$ filters), a few times deeper than the actual
on-axis GRB rate, indicating that LSST is likely to detect such
events.

The LSST data set can form the basis for a search similar to the one
described here, with the main difference that contemporaneous color
information may not be available.  However, with a ${\rm S/N}\approx
10$ cut as imposed here, the observations will reach about 1 mag
deeper than our search, indicating that they will uncover M dwarf
flares to distances that are only about $50\%$ larger than those found
here (i.e., generally $\lesssim 2$ kpc).  At the same time, the LSST
templates will reach a comparable depth to the PS1/MDS templates after
only $\sim 10$ visits per field per filter, i.e., within a year of
when the survey commences.  Thus, we conclude that it will be simple
to identify M dwarf counterparts in the template images for
essentially all M dwarf flares, negating the need for follow-up
spectroscopy.  This approach will effectively eliminate the largest
known contaminating foreground for $\sim{\rm hour}$ timescales.

Similarly, the time baseline of $15-60$ min per field will allow for
astrometric rejection of most asteroids even near the stationary
point, although as we found in our PS1/MDS search some asteroids have
negligible motions that are indistinguishable from the astrometric
scatter of field sources ($2/8$ in our search).  Such asteroids can in
principle be recognized using color information, but this is not
likely to be available with LSST.  On the other hand, a constant
brightness level between visits $15-60$ min apart, combined with a
location near the ecliptic plane and in particular with solar
elongation of $\sim 130$ deg, will be indicative of an asteroid
origin.  Such fields can simply be avoided in searches for fast
transients.

Thus, the two primary contaminants for extragalactic fast optical
transient searches (M dwarf flares and asteroids near the stationary
point) will be identifiable with the LSST baseline survey strategy.
Furthermore, we anticipate based on the comparison to known transient
classes that with a year of operations LSST is unlikely to reveal
large numbers of extragalactic fast transients unless they result from
a source population that far exceeds classical novae, supernova
progenitors, or compact object binaries (Figure~\ref{fig:surveys2}).

\section{Conclusions}
\label{sec:conc}

We present a search for fast optical transients on a timescale of
about 0.5 hr to 1 d in consecutive \gps\rps\ observations of the
PS1/MDS fields, by requiring a detection in both filters with no
additional detections on preceding or subsequent nights.  The search
yielded 19 astrophysical transients, of which 8 events that lack
quiescent counterparts are identified as main-belt asteroids near the
stationary point of their orbits, while the remaining 11 transients
are identified as flares from M5--M9 dwarf stars at distances of about
$0.2-1.2$ kpc.  The flare properties are generally similar to those
from M dwarfs studied in the SDSS Stripe 82, although our sample
extends to later spectral types and to much greater vertical distances
from the Galactic plane.

A key result of our search is a limit on the sky-projected rate of
extragalactic fast transients of $R_{\rm FOT}\lesssim 0.12$ deg$^{-2}$
d$^{-1}$ ($\sim 0.5$ hr) that is about $30-50$ time deeper than
previous limits; the limit for a timescale of $\sim {\rm day}$ is
$\lesssim 2.4\times 10^{-3}$ deg$^{-2}$ d$^{-1}$.  The upper bounds on
the volumetric rates at fiducial absolute magnitudes of $-10$, $-14$,
and $-24$ mag are likewise an order of magnitude deeper than from
previous searches (Table~\ref{tab:rates}).  With an additional $3$
years of PS1/MDS data in hand we can improve these estimates by a
factor of a few.  We also anticipate additional M dwarf flare
detections that will allow us to better characterize the distribution
of flare properties, as well as the properties of flaring M dwarfs in
general.

To guide the conversion from our sky-projected rate to volumetric
rates, and to motivate future searches for fast transients, we also
explore the expected luminosities of such transients for a range of
physically motivated models.  We find that non-relativistic fast
transients are generally limited to about $-10$ mag for a timescale of
$\sim 0.5$ hr and $-14$ mag for a timescale of $\sim {\rm day}$ even
if powered by radioactive decay.  This is simply a reflection of the
low ejecta mass required to achieve a rapid diffusion timescale of
$\lesssim{\rm day}$.  Such low luminosity events are best explored
through targeted searches of galaxies in the local universe, with
anticipated event rates of $\lesssim 10^{-4}$ Mpc$^{-3}$ yr$^{-1}$
(Figure~\ref{fig:surveys2}).  A separate class of relativistic fast
transients (on-axis GRBs, magnetar engines) can exceed $-25$ mag on a
timescale of $\sim 0.5$ hr, but they are exceedingly rare, with
anticipated volumetric rates of $\sim 10^{-8}$ Mpc$^{-3}$ yr$^{-1}$
(or sky-projected rates of a few per sky per day).  Such event rates
are orders of magnitude below the level probed in current searches,
but will be reached by the LSST main survey in a full year.  Another
strategy to find such events is a shallower search of a wider sky area
than is planned for LSST.  We note that a broad exploration of the
various mechanisms that can power optical transients is essential to
guide and motivate future searches for fast extragalactic transients.
The initial investigation performed here already places clear bounds
on the luminosities and rates for a wide range of mechanisms.

Finally, since the PS1/MDS survey is currently the only close analogue
to the main LSST survey in terms of depth, cadence, and choice of
filters, we use the results of our search to investigate the efficacy
of fast transient searches with LSST.  We demonstrate that the main
contaminants recognized here should be identifiable with the LSST
survey strategy, without the need for expensive follow-up
spectroscopy.  Namely, asteroids near the stationary point can be
recognized through a larger than average astrometric shift, a constant
brightness in visits separated by $15-60$ min, and an expected
location near the ecliptic plane with solar elongation of $130$ deg.
LSST may not provide near-simultaneous color information, which can
serve as an additional discriminant for asteroids.  For M dwarf flares
we show that LSST's increased sensitivity will probe a larger volume
of the Galaxy, but the correspondingly deeper templates will still
allow for the identification of quiescent counterparts in essentially
all cases.  The PS1/MDS survey demonstrates that with multi-band
photometry it is possible to identify M dwarf flares without the need
for follow-up spectroscopy (as in previous searches; e.g.,
\citealt{kr06}).  This result indicates that extragalactic fast
optical transients should be able to pierce the veil of foreground
flares.

\acknowledgments We thank Mario Juric for helpful information on the
planned LSST survey strategy.  E.~B.~acknowledges support for this
work from the National Science Foundation through Grant AST-1008361.
PS1 has been made possible through contributions of the Institute for
Astronomy, the University of Hawaii, the Pan-STARRS1 Project Office,
the Max-Planck Society and its participating institutes, the Max
Planck Institute for Astronomy, Heidelberg and the Max Planck
Institute for Extraterrestrial Physics, Garching, The Johns Hopkins
University, Durham University, the University of Edinburgh, Queen's
University Belfast, the Harvard-Smithsonian Center for Astrophysics,
and the Las Cumbres Observatory Global Telescope Network,
Incorporated, the National Central University of Taiwan, and the
National Aeronautics and Space Administration under Grant NNX08AR22G
issued through the Planetary Science Division of the NASA Science
Mission Directorate.

{\it Facilities:} \facility{Pan-STARRS1}


\clearpage
\begin{deluxetable*}{lccclllcc}
\tabletypesize{\scriptsize}
\tablecolumns{9} 
\tablewidth{0pt} 
\tablecaption{Fast Optical Transients with Quiescent Counterparts 
\label{tab:mdwarfs}}
\tablehead{
\colhead{OBJID}                    &                                   
\colhead{R.A.}                     &
\colhead{Decl.}                    &
\colhead{UT Date}                  &
\colhead{$m_{\rm flare,P1}$}       &                             
\colhead{$m_{\rm q,P1}$}           &                             
\colhead{$m_{\rm q,SDSS}$}         &                              
\colhead{Sp.~T.}                   &                              
\colhead{$d$}                      \\                              
\colhead{}                         &        
\colhead{($^{\rm h}\, ^{\rm m}\, ^{\rm s}$)} &
\colhead{($^\circ\,'\,''$)}        &
\colhead{}                         &               
\colhead{(AB mag)}                 &            
\colhead{(AB mag)}                 &            
\colhead{(AB mag)}                 &                                         
\colhead{}                         &               
\colhead{(pc)}                                   
}
\startdata
PSO\,J52.0080$-$27.0515 & \ra{03}{28}{01.929} & \dec{$-$27}{03}{05.51} & 2010-09-07 & $g=22.14\pm 0.10$ & $g=21.55\pm 0.05$ & \nod & M5 & 315 \\
& & & & $r=21.23\pm 0.10$ & $r=20.01\pm 0.03$ & \nod &    &     \\
& & & & & $i=18.35\pm 0.03$ & \nod &    &     \\
& & & & & $z=17.46\pm 0.04$ & \nod &    &     \\\hline
PSO\,J130.6460$+$45.2130 & \ra{08}{42}{35.063} & \dec{+45}{12}{47.07} & 2009-12-14 & $g=22.07\pm 0.05$ & $g\gtrsim 25.28$  & \nod & M9 & 290 \\
& & & & $r=22.05\pm 0.05$ & $r\gtrsim 25.31$  & \nod &    &     \\
& & & & & $i=22.56\pm 0.04$ & \nod &    &     \\
& & & & & $z=21.00\pm 0.02$ & \nod &    &     \\\hline
PSO\,J162.3497$+$56.6609 & \ra{10}{49}{23.934} & \dec{+56}{39}{39.57} & 2011-04-29 & $g=18.48\pm 0.02$ & $g=23.62\pm 0.07$ & $g=23.45\pm 0.23$ & M7.5 & 220 \\
& & & & $r=18.93\pm 0.05$ & $r=22.19\pm 0.04$ & $r=22.35\pm 0.16$ &      &     \\
& & & & & $i=19.96\pm 0.01$ & $i=19.97\pm 0.03$ &      &     \\
& & & & & $z=18.78\pm 0.04$ & $z=18.69\pm 0.04$ &      &     \\\hline 
PSO\,J164.3814$+$58.3011 & \ra{10}{57}{31.557} & \dec{+58}{18}{04.12} & 2011-04-14 & $g=19.08\pm 0.01$ & $g\gtrsim 25.25$  & \nod & M8.5 & 330 \\
& & & & $r=21.08\pm 0.04$ & $r=24.07\pm 0.10$ & $r=23.51\pm 0.31$ &    &     \\
& & & & & $i=21.72\pm 0.02$ & $i=21.45\pm 0.08$ &    &     \\
& & & & & $z=20.20\pm 0.01$ & $z=19.79\pm 0.07$ &    &     \\\hline
PSO\,J184.0215$+$47.4267 & \ra{12}{16}{05.180} & \dec{+47}{25}{36.24} & 2010-06-06 & $g=20.61\pm 0.03$ & $g=21.88\pm 0.05$ & $g=21.83\pm 0.06$ & M4.5 & 600 \\
& & & & $r=21.10\pm 0.10$ & $r=20.47\pm 0.03$ & $r=20.40\pm 0.03$ &      &     \\
& & & & & $i=18.96\pm 0.04$ & $i=18.87\pm 0.01$ &      &     \\
& & & & & $z=18.14\pm 0.05$ & $z=18.05\pm 0.02$ &      &     \\\hline
PSO\,J184.6519$+$48.1834 & \ra{12}{18}{36.472} & \dec{+48}{11}{00.41} & 2010-02-09 & $g=18.96\pm 0.01$ & $g=20.12\pm 0.02$ & $g=20.18\pm 0.02$ & M4.5 & 280 \\
& & & & $r=19.60\pm 0.02$ & $r=18.80\pm 0.02$ & $r=18.75\pm 0.01$ &      &     \\
& & & & & $i=17.32\pm 0.04$ & $i=17.34\pm 0.01$ &      &     \\
& & & & & $z=16.66\pm 0.04$ & $z=16.57\pm 0.01$ &      &     \\\hline 
PSO\,J185.4857$+$47.2586 & \ra{12}{21}{56.583} & \dec{+47}{15}{31.14} & 2011-06-07 & $g=22.21\pm 0.10$ & $g\gtrsim 25.10$  & \nod & M8 & 1160 \\
& & & & $r=22.33\pm 0.10$ & $r\gtrsim 24.95$  & \nod &    &     \\
& & & & & $i=24.25\pm 0.20$ & \nod &    &     \\
& & & & & $z=22.85\pm 0.11$ & \nod &    &     \\\hline
PSO\,J211.8195$+$52.9681 & \ra{14}{07}{16.681} & \dec{+52}{58}{05.27} & 2011-06-07 & $g=20.60\pm 0.02$ & $g=21.95\pm 0.04$ & $g=21.92\pm 0.08$ & M4.5 & 650 \\
& & & & $r=20.62\pm 0.02$ & $r=20.63\pm 0.02$ & $r=20.57\pm 0.04$ &      &     \\
& & & & & $i=19.15\pm 0.02$ & $i=19.11\pm 0.02$ &      &     \\
& & & & & $z=18.39\pm 0.02$ & $z=18.27\pm 0.03$ &      &     \\\hline
PSO\,J214.7904$+$53.6886 & \ra{14}{19}{09.715} & \dec{+53}{41}{18.99} & 2010-02-06 & $g=21.60\pm 0.08$ & $g=20.11\pm 0.02$ & $g=20.22\pm 0.02$ & M5 & 220 \\
& & & & $r=20.86\pm 0.08$ & $r=18.71\pm 0.02$ & $r=18.71\pm 0.01$ &    &     \\
& & & & & $i=17.10\pm 0.02$ & $i=17.10\pm 0.01$ &    &     \\
& & & & & $z=16.38\pm 0.04$ & $z=16.23\pm 0.01$ &    &     \\\hline
PSO\,J334.5602$-$0.8479 & \ra{22}{18}{14.455} & \dec{-00}{50}{52.77} & 2011-07-31 & $g=22.45\pm 0.08$ & $g=22.67\pm 0.04$ & $g=22.78\pm 0.17$ & M4.5 & 690 \\
& & & & $r=22.40\pm 0.10$ & $r=21.23\pm 0.02$ & $r=21.20\pm 0.05$ &      &     \\
& & & & & $i=19.67\pm 0.02$ & $i=19.66\pm 0.02$ &      &     \\
& & & & & $z=18.84\pm 0.02$ & $z=18.87\pm 0.04$ &      &     \\\hline
PSO\,J334.7536$+$0.8769 & \ra{22}{19}{00.865} & \dec{+00}{52}{37.11} & 2010-09-04 & $g=21.93\pm 0.06$ & $g=23.32\pm 0.06$ & $g=22.94\pm 0.20$ & M5.5 & 570 \\
& & & & $r=21.84\pm 0.06$ & $r=21.77\pm 0.02$ & $r=21.76\pm 0.09$ &      &     \\
& & & & & $i=20.02\pm 0.01$ & $i=19.99\pm 0.03$ &      &     \\
& & & & & $z=19.12\pm 0.01$ & $z=19.09\pm 0.06$ &      &      
\enddata
\tablecomments{Properties of the fast transients that exhibit
quiescent counterparts, as well as multi-band photometry of the
counterparts.  We find that all 11 transients arise from M dwarfs, and
provide the inferred spectral types and distances of these sources.}
\end{deluxetable*}

\begin{deluxetable*}{lccccccc}
\tabletypesize{\scriptsize}
\tablecolumns{8} 
\tablewidth{0pt} 
\tablecaption{Fast Optical Transients Lacking Quiescent Counterparts 
\label{tab:asteroids}}
\tablehead{
\colhead{OBJID}         &                                   
\colhead{R.A.}          &
\colhead{Decl.}         &
\colhead{$\ell$}        &
\colhead{$b$}           &
\colhead{UT Date}       &
\colhead{$g_{\rm P1}$}  &                             
\colhead{$r_{\rm P1}$}  \\                              
\colhead{}              &        
\colhead{($^{\rm h}\, ^{\rm m}\, ^{\rm s}$)} &
\colhead{($^\circ\,'\,''$)}                  &
\colhead{(deg)}         &               
\colhead{(deg)}         &               
\colhead{}              &        
\colhead{(AB mag)}      &            
\colhead{(AB mag)}                             
}
\startdata
PSO\,J149.0662$+$1.8129 & \ra{09}{56}{15.886} & \dec{+01}{48}{46.46}   & $150.5546$ & $-10.0989$ & 2011-04-23 & $22.73\pm 0.08$ & $22.14\pm 0.06$ \\
PSO\,J149.2788$+$3.3718 & \ra{09}{57}{06.910} & \dec{+03}{22}{18.37}   & $150.2096$ & $-8.5629$ & 2011-04-23 & $19.78\pm 0.02$ & $19.25\pm 0.03$ \\
PSO\,J149.3234$+$2.6863 & \ra{09}{57}{17.622} & \dec{+02}{41}{10.51}   & $150.4922$ & $-9.1906$ & 2011-04-23 & $20.44\pm 0.03$ & $19.90\pm 0.03$ \\
PSO\,J333.1738$+$0.4535 & \ra{22}{12}{41.723} & \dec{+00}{27}{12.69}   & $335.2762$ & $+10.7642$ & 2011-10-24 & $22.13\pm 0.06$ & $21.47\pm 0.04$ \\
PSO\,J333.9157$-$0.6311 & \ra{22}{15}{39.776} & \dec{$-$00}{37}{51.89} & $335.5803$ & $+9.4847$ & 2010-10-10 & $21.53\pm 0.06$ & $20.96\pm 0.03$ \\
PSO\,J352.2672$+$1.2637 & \ra{23}{29}{04.139} & \dec{$-$01}{15}{49.35} & $352.3995$ & $+1.9068$ & 2010-11-03 & $21.35\pm 0.04$ & $20.55\pm 0.02$ \\
PSO\,J352.5520$+$0.8004 & \ra{23}{30}{12.469} & \dec{+00}{48}{01.49}   & $353.4772$ & $+3.6909$ & 2010-11-03 & $22.40\pm 0.10$ & $21.67\pm 0.07$ \\
PSO\,J352.5968$-$0.4471 & \ra{23}{30}{23.229} & \dec{$-$00}{26}{49.64} & $353.0249$ & $+2.5271$ & 2010-10-31 & $19.59\pm 0.02$ & $18.79\pm 0.02$
\enddata
\tablecomments{Properties of the fast transients that lack quiescent
counterparts.  We find that all 8 transients exhibit the colors and
expected ecliptic coordinates of main-belt asteroids near the
stationary point of their orbits.}
\end{deluxetable*}

\begin{deluxetable*}{lcclllll}
\tabletypesize{\footnotesize}
\tablecolumns{8} 
\tablewidth{0pt} 
\tablecaption{Surveys for Fast Optical Transients with a Timescale of
$\sim 0.5$ hr \label{tab:rates}}
\tablehead{
\colhead{Survey}                        &                                   
\colhead{Areal Exposure}                &                                   
\colhead{Limiting Magnitude}            &
\colhead{$R_{\rm FOT}$}                 &
\colhead{$R_{\rm FOT}(-10\,{\rm mag})$} &
\colhead{$R_{\rm FOT}(-14\,{\rm mag})$} &
\colhead{$R_{\rm FOT}(-24\,{\rm mag})$} &
\colhead{References}                    \\
\colhead{}                              &
\colhead{(deg$^2$ d)}                   &                             
\colhead{}                              &     
\colhead{(deg$^{-2}$ d$^{-1}$)}         &        
\colhead{(Mpc$^{-3}$ yr$^{-1}$)}        &        
\colhead{(Mpc$^{-3}$ yr$^{-1}$)}        &
\colhead{(Mpc$^{-3}$ yr$^{-1}$)}        &
\colhead{}                                   
}
\startdata
PS1/MDS   & 40.4 & 22.5 & $\lesssim 0.12$  & $\lesssim 13$             & $\lesssim 0.05$ & $\lesssim 1\times 10^{-6}$ & This paper     \\
DLS       & 1.1  & 23.8 & $\lesssim 6.5$   & $\lesssim 1.3\times 10^2$ & $\lesssim 0.5$  & $\lesssim 6\times 10^{-5}$ & \citet{bwb+04} \\
Fornax    & 1.9  & 21.3 & $\lesssim 3.3$   & $\lesssim 2.0\times 10^3$ & $\lesssim 7.8$  & $\lesssim 3\times 10^{-5}$ & \citet{rok+08} \\
ROTSE-III & 635  & 17.5 & $\lesssim 0.005$ & \nod                      & \nod            & $\lesssim 6\times 10^{-6}$ & \citet{raa+05} \\
MASTER    & \nod & 17.5 & $\lesssim 0.003$ & \nod                      & \nod            & $\lesssim 4\times 10^{-6}$ & \citet{lkk+07} \\\hline
LSST      & 62   & 23.5 & $\lesssim 0.07$  & $\lesssim 2$              & $\lesssim 8\times 10^{-3}$ & $\lesssim 5\times 10^{-7}$ & \nod           \\
          & $1.6\times 10^4$ & 23.5 & $\lesssim 3\times 10^{-4}$ & $\lesssim 9\times 10^{-3}$ & $\lesssim 3\times 10^{-5}$ & $\lesssim 2\times 10^{-9}$ &\nod
\enddata
\tablecomments{Survey parameters and resulting limits on the projected
sky rates and volumetric rates of extragalactic fast optical
transients with a timescale of $\sim 0.5$ hr.}
\end{deluxetable*}

\end{document}